\documentclass[prb,aps,twocolumn,showpacs]{revtex4}

\usepackage{epsfig}
\usepackage{natbib}

\begin{document}

\title{Phase separation, competition, and volume-fraction control in
NaFe$_{1-x}$Co$_x$As}

\author{Long Ma$^{1,2}$}
\author{J. Dai$^{1}$}
\author{P. S. Wang$^{1}$}
\author{X. R. Lu$^{1}$}
\author{Yu Song$^{3,4}$}
\author{Chenglin Zhang$^{3,4}$}
\author{G. T. Tan$^{4,5}$}
\author{Pengcheng Dai$^{3}$}
\author{D. Hu$^{6}$}
\author{S. L. Li$^{6,7}$}
\author{B. Normand$^{1}$}
\author{Weiqiang Yu$^{1}$}
\email{wqyu_phy@ruc.edu.cn}
\affiliation{$^{1}$Department of Physics, Renmin University of China, Beijing
100872, China \\
$^{2}$High Magnetic Field Laboratory, Chinese Academy of Sciences, Hefei 
230031, China\\
$^{3}$Department of Physics and Astronomy, Rice University, Houston,
Texas 77005-1827, USA \\
$^{4}$Department of Physics and Astronomy, University of Tennessee,
Knoxville, Tennessee 37996-1200, USA \\
$^{5}$Department of Physics, Beijing Normal University, Beijing 100875, China\\
$^{6}$Beijing National Laboratory for Condensed Matter Physics, Institute of
Physics, Chinese Academy of Sciences, Beijing 100190, China \\
$^{7}$Collaborative Innovation Center of Quantum Matter, Beijing, China}
\date{\today}


\begin{abstract}

We report a detailed nuclear magnetic resonance (NMR) study by combined 
$^{23}$Na and $^{75}$As measurements over a broad range of doping to map 
the phase diagram of NaFe$_{1-x}$Co$_x$As. In the underdoped regime ($x 
\le$ 0.017), we find a magnetic phase with robust antiferromagnetic (AFM) 
order, which we denote the {\it s}-AFM phase, cohabiting with a phase of 
weak and possibly proximity-induced AFM order ({\it w}-AFM) whose volume 
fraction $V \simeq 8$\% is approximately constant. Near optimal doping, at 
$x = 0.0175$, we observe a phase separation between static antiferromagnetism 
related to the {\it s}-AFM phase and a paramagnetic (PM) phase related to 
{\it w}-AFM. The volume fraction of AFM phase increases upon cooling, but 
both the N{\'e}el temperature and the volume fraction can be suppressed 
systematically by applying a $c$-axis magnetic field. On cooling below 
$T_c$, superconductivity occupies the PM region and its volume fraction 
grows at the expense of the AFM phase, demonstrating a phase separation 
of the two types of order based on volume exclusion. At higher dopings, 
static antiferromagnetism and even critical AFM fluctuations are completely 
suppressed by superconductivity. Thus the phase diagram we establish contains 
two distinct types of phase separation and reflects a strong competition 
between AFM and superconducting phases both in real space and in momentum 
space. We suggest that both this strict mutual exclusion and the robustness 
of superconductivity against magnetism are consequences of the extreme 
two-dimensionality of NaFeAs.

\end{abstract}

\maketitle

\section{Introduction}

Competing electronic phases underlie a number of the most unconventional
phenomena in condensed matter systems. When this competition is sufficiently
strong, the usual outcome is a phase separation. One outstanding example of
this situation is provided by materials showing colossal magnetoresistance,
where competing magnetic interactions lead to phase separation between
conducting ferromagnetic and insulating antiferromagnetic (AFM) regions
\cite{Cheong_Nature_1999}. As a consequence, an external magnetic field
can be used to control the resistance over many orders of magnitude,
offering possible applications in electronic devices. In cuprate
superconductors, the competition between antiferromagnetism and
superconductivity forms the basis for the majority of the observed
phenomena and for several classes of materials the debate can be phrased
in terms of the extent to which phase separation is the outcome. The stripe
phase, which has been the object of heated research interest for two decades,
can be considered as a form of atomic-scale phase separation between AFM and 
superconducting (SC) regions, and such self-organizing heterostructures are 
a direct reflection of the electronic correlations whose effects are essential 
to understanding the mechanism of high-temperature superconductivity 
\cite{Kivelson_RMP_75_1201}.

The competition between AFM and SC phases also forms the foundation for 
the physics of iron-based superconductors \cite{Hosono_Jacs_130_3296,
Chen_PRL_100_247002, Chen_Nature_453_761, Ren_CPL_12_105, Cruz_Nat_453_899},
where it is manifest in the emergence of a tetragonal SC phase upon doping-
or pressure-induced suppression of an orthorhombic AFM phase. Iron-based 
superconductors have in common a quasi-two-dimensional atomic structure of 
weakly coupled FeAs or FeSe planes, although the exact crystal structure 
varies somewhat among the 1111, 122, 111, and 11 families of materials 
\cite{Paglione_Nature}; in fact the phenomena we report here will highlight 
some of the important differences arising between families as a consequence 
of the strength of their interplane coupling. Currently, the detailed phase 
diagram close to optimal doping (the concentration giving the maximum SC 
transition temperature, $T_c$) remains hotly debated, with evidence cited 
in favor of phase coexistence, of a possible AFM quantum critical point, or 
of heterostructures of AFM and SC phases. The levels of doping and disorder, 
and their impact on the phases and their competition, seem to vary between 
structural families, defying any search for universal properties. However, 
this variety does open additional avenues in the search for novel forms of 
phase separation or heterostructure formation, and with them the scope for 
obtaining further clues to the mechanism of high-$T_c$ superconductivity.

The 111 family is based on the materials LiFeAs and NaFeAs, with doping 
effected most easily as NaFe$_{1-x}$Co$_x$As. The parent compound NaFeAs has 
a separate structural transition ($T_s \simeq$ 55 K) and magnetic transition 
($T_N \simeq$ 41 K) \cite{Chen_PRL_102, SLPRB2009}, the latter to an AFM phase 
with small ordered moments ($\mu \simeq$ 0.32$\mu_B$/Fe) \cite{Ma_NaFeAs}. 
Here we choose to use the notation $T_N$, rather than $T_{SDW}$, to reflect 
the strong local-moment character of the magnetic phase, an issue to which 
we return in Sec.~VI. The separation of $T_s$ and $T_N$, and also the 
relatively low $T_N$ values, count among the initial pieces of evidence for 
a rather weak interlayer coupling \cite{XuCK} between FeAs planes in the 
NaFeAs system. The crystal quality, particularly the homogeneity of dopant 
distribution, is thought to be among the best in any iron-based 
superconductors, as measured in transition widths and observed by scanning 
tunneling microscopy (STM). Nevertheless, for underdoped 111 compounds a 
coexistence of inhomogeneous antiferromagnetism and superconductivity 
has been suggested by transport \cite{Chen_NaFeCoAs}, angle-resolved 
photoemission spectroscopy (ARPES) \cite{Feng_PRX_2013}, and STM measurements 
\cite{WangYY_NC_2012}. By contrast, the ``coexistence'' of a strongly ordered 
AFM phase ({\it s}-AFM) and a weakly ordered one ({\it w}-AFM) on different 
spatial sites (this situation may be denoted more specifically as a 
``cohabitation'') has been reported from NMR measurements \cite{Halperin_2013}. 
Clearly a phase inhomogeneity is observed in all of these studies. However, a 
detailed analysis of the intrinsic properties of the primary phases, of the 
exact phase diagram around optimal doping, and of the different phase volume 
fractions, is still required.

In this paper, we exploit the power of NMR as a completely local probe to
resolve the appearance and properties of the different AFM and SC phases
in NaFe$_{1-x}$Co$_x$As. For low dopings, we confirm the cohabitation of
two regimes, {\it s}-AFM and {\it w}-AFM, finding that the {\it w}-AFM
phase has a constant volume fraction of order $8\%$; this indicates an
intrinsic effect unrelated to the doping concentration and we suggest
that the {\it w}-AFM phenomenon is actually a proximity-induced moment
distribution in a paramagnetic (PM) phase. For dopings around optimal,
we find at $x \simeq$ 0.0175 ($T_c \simeq$ 20 K) the onset of regions
of antiferromagnetism below 25 K, where the AFM volume grows with cooling 
but the application of a magnetic field suppresses both $T_N$ and the 
magnetic volume fraction. At lower temperatures, superconductivity enters 
in the PM phase and its volume fraction increases at the expense of the 
AFM region both on cooling and (somewhat paradoxically) with increasing 
field. For $x \simeq 0.019$ ($T_c \simeq$ 22 K), superconductivity 
suppresses not only AFM order but also the critical AFM fluctuations 
below $T_c$, forming in the terminology of some authors the mechanism 
by which the AFM quantum critical point is ``avoided.''

Our results present direct evidence for the mutual exclusion of 
antiferromagnetism and superconductivity, which leads to a ``volume 
competition'' between regions of established (finite-order-parameter) 
phases replacing each other in space in a first-order manner. This volume 
competition can be controlled systematically by the temperature and magnetic 
field, and we suggest that it exists in many other iron-based superconductors. 
Our detailed studies of the spin-lattice relaxation rate across the phase 
diagram indicate the importance of both itinerant (conduction-electron) and 
local-moment (valence-electron) contributions to both types of order. A
theoretical interpretation of the strong competition points to the key role
of the very two-dimensional (2D) Fermi surfaces in 111 systems and to
orbital-selective phenomena depending on the specific bands involved at
the different Fermi surfaces.

The structure of this paper is as follows. In Sec.~II we summarize our
basic sample properties and measurement procedures. We begin the presentation
of our results in Sec.~III by considering the nature of the AFM phases in the
underdoped regime. In Sec.~IV we focus on our samples close to optimal doping
to elucidate the nature of phase cohabitation and volume competition. With
these results in hand, in Sec.~V we complete a detailed phase diagram for
the NaFe$_{1-x}$Co$_x$As system. Section VI contains an interpretation of our
results and a discussion of their implications for the understanding of
superconductivity in iron-based materials, concluded by a short summary.

\section{Materials and Methods}

Our NaFe$_{1-x}$Co$_x$As single crystals are synthesized by the flux-grown
method with NaAs as the flux \cite{LiSL_PRB_2013}. The Co doping levels are
monitored by the inductively coupled plasma (ICP) technique. However, ICP
measurements are subject to significant inaccuracies and are by no means
appropriate to establish the doping concentrations to the degree of precision
required to study the NaFe$_{1-x}$Co$_x$As system, where all doping levels are
anomalously low (optimal doping $x_{\rm opt} = 1.9$\%). Here we report the 
nominal stoichiometries of the different crystals and establish their 
relative doping values from our physical measurements by seeking continuity
and possible irregularities (Sec.~V).

The SC transition temperature $T_c$ was determined {\it in situ} by the 
sudden decrease in inductance of the NMR coil. The zero-field $T_c$ values 
at different dopings are consistent with earlier reports \cite{LiSL_PRB_2013}. 
We have performed NMR measurements on both the $^{23}$Na and $^{75}$As nuclei, 
with the field applied both within the crystalline $ab$-plane and along the 
$c$-axis. We use a TecMag spectrometer and obtain the NMR spectra from the 
Fourier transform of the spin-echo signal. The spin-lattice relaxation rates 
$1/^{23}T_1$ and $1/^{75}T_1$ were measured by the inversion method and all 
magnetization recovery rates could be fitted well with the function $1 - I(t)
/I_0 = A (0.1 e^{-t/T_1} + 0.9 e^{-6t/T_1})$ (appropriate for $I = 3/2$ nuclei).

For detecting the magnitude of the different ordered magnetic moments in 
the NaFe$_{1-x}$Co$_x$As system, it is important to be able to use both the 
$^{23}$Na and $^{75}$As spectra, in order to exploit their very different 
relative hyperfine coupling strengths, $^{75}A_{hf}/^{23}A_{hf} \simeq 23$ 
\cite{Ma_NaFeAs}. Because $^{23}$Na has a rather weak hyperfine coupling, 
the resulting narrow line width makes it a very accurate probe of 
magnetically ordered states with inhomogeneous moment distributions. By 
contrast, $^{75}$As has a strong hyperfine coupling, making it sensitive 
to very weak ordered moments and ideal for proving the absence of magnetic 
order in a true PM phase.

\begin{figure}
\includegraphics[width=8.5cm, height=8cm]{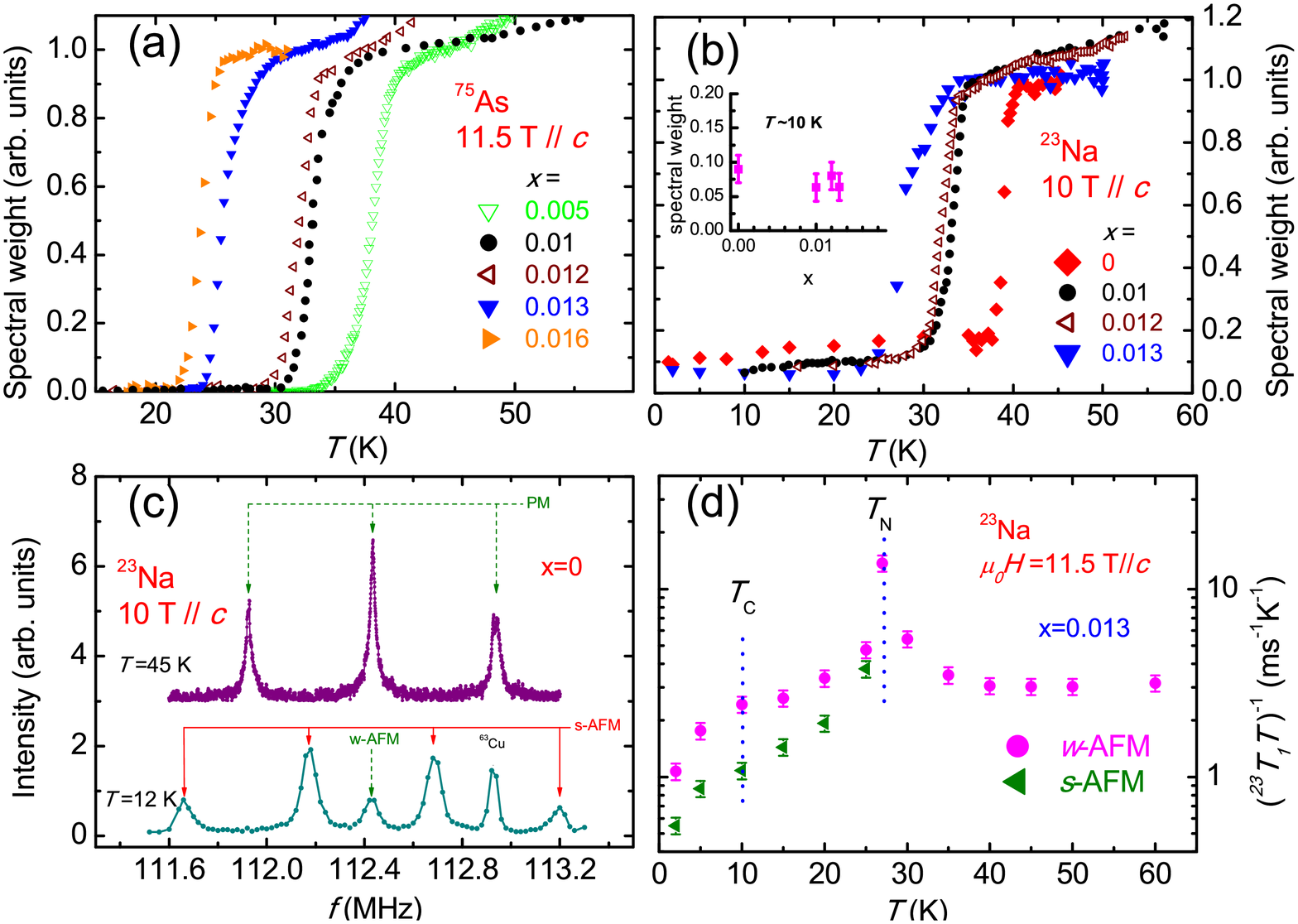}
\caption{\label{tn1}(Color online) (a) Spectral weight of the $^{75}$As
PM signals of NaFe$_{1-x}$Co$_x$As samples with five different dopings $x$, 
shown as a function of temperature and measured with a field of 11.5 T 
applied along the $c$-axis. The sharp loss of spectral weight indicates 
the onset of AFM ordering. (b) Spectral weight of the $^{23}$Na signal at 
the center frequency as a function of temperature for four sample dopings. 
The inset shows the residual spectral weight of the center line at low 
temperatures. (c) $^{23}$Na spectra for the parent compound ($x = 0$) at
temperatures above and far below $T_N$. (d) Spin-lattice relaxation rate
$1/^{23}T_1T$ for $x =$ 0.013; dotted lines mark the onset temperatures
$T_N$ and $T_c$ for AFM and SC order at the measurement field of 11.5 T.}
\end{figure}

\section{Nature of Underdoped Phase Separation}

We begin by considering our underdoped samples to investigate the
potentially inhomogeneous AFM phases reported previously \cite{Halperin_2013}.
The onset temperature for the transition to magnetic order can be determined
from the $^{75}$As and $^{23}$Na NMR spectra. Figure \ref{tn1}(a) shows the
spectral weight of the center peak of the $^{75}$As spectrum as a function 
of temperature for five underdoped samples. This quantity is the PM signal 
and it drops sharply at the onset of magnetic order, as the character of 
the magnetic environment is altered and spectral weight is transferred
away from the center. We determine the N\'eel temperature $T_N$ at each
doping and we note that, far below $T_N$, the spectral weight appears to
decrease to zero, indicating no residual PM phase at any doping. In
Fig.~\ref{tn1}(b) we show the spectral weight of $^{23}$Na at the center
frequency for four dopings up to $x = 0.013$. Again the sharp drop of
spectral weight indicates the onset of AFM order. However, a residual 8$\%$
spectral weight persists far below $T_N$ for all dopings below $x = 0.017$
[inset, Fig.~\ref{tn1}(b)]. The spectra for the parent phase ($x = 0$),
shown in Fig.~\ref{tn1}(c), contain a PM signal with one center peak and
two satellites above $T_N$. Far below $T_N$, there is a clear spectral
splitting due to strong magnetic order, accompanied by a residual peak
in the center. This result provides a good example of the sensitivity of
$^{23}$Na measurements: our data demonstrate that a very weak AFM order 
must be present to account for the residual spectral weight and line-width
broadening. This is consistent with the absence of the $^{75}$As PM signal,
although the strong hyperfine coupling of $^{75}$As makes it difficult to
discern the nature of the magnetic state. Below $T_N$, the line width of
the $^{23}$Na spectrum is approximately 35 kHz, and therefore the upper
bound on the ordered moments is only $6.5\%$ of that in the parent
compound, where the $^{23}$Na spectrum is split by 540 kHz.

The NMR study of Oh {\it et al.} \cite{Halperin_2013} reports two species
of antiferromagnetism in a sample with $x \approx$ 0.017, one with a large
ordered moment ({\it s}-AFM) and the other one with a small moment
({\it w}-AFM). Our data from lower dopings are consistent with the finding
of a small volume fraction of a {\it w}-AFM phase \cite{Halperin_2013}, but
we find [inset, Fig,~\ref{tn1}(b)] that this volume fraction does not change
with doping. This result indicates that the appearance of the {\it w}-AFM
phase is not an intrinsic consequence of Co doping; if it is a disorder
effect then it must be of a different type, perhaps with its origin in a
strain or chemical inhomogeneity. In Fig.~\ref{tn1}(d), we present the
spin-lattice relaxation rate, $1/^{23}T_1T$, measured at the peak frequencies
of both the {\it s}-AFM and the {\it w}-AFM signals. As noted in Sec.~II the 
magnetization recoveries at the two separate frequencies each follow the 
single-component function expected for $I = 3/2$ nuclei. The relaxation 
rates for both signals fall at the same ordering temperature $T_N$. For the 
{\it w}-AFM component, a phase having such a small ordered moment but a high 
onset temperature $T_N$ is generically very unlikely, and we suggest that 
the consistent explanation is a microscopic phase separation into AFM and PM 
regimes, but with weak magnetic order (appearing as the {\it w}-AFM phase) 
induced in the PM phase by its proximity to the {\it s}-AFM one. We note 
in addition that the significantly faster $^{23}T_1$ below $T_N$ in the 
{\it w}-AFM phase [Fig.~\ref{tn1}(d)] is also consistent with spin 
fluctuations being only partially suppressed by a weak proximity effect.

The concept of ``nanoscale phase separation'' in iron-based superconductors 
is known from the depleted iron selenide materials $A_2$Fe$_4$Se$_5$ (A = K, 
Rb, Cs, Tl; ``245'') \cite{rme}, which appear to show a robust AFM phase 
accompanied by an equally robust but quite separate PM phase; the latter is 
the only part of the system to turn SC at $T_c$, forming a percolating SC phase 
despite having a volume fraction below 10\%. However, it is generally thought 
that this phase separation is primarily a consequence of vacancy-induced 
structural inhomogeneity, causing a clear doping inhomogeneity, whereas our 
results (previous paragraph) appear to exclude this in NaFe$_{1-x}$Co$_x$As. 
Here we observe for $x = 0.013$ [Fig.~\ref{tn1}(d)] that $1/^{23}T_1T$ shows 
a similar fall in both the {\it s}- and {\it w}-AFM signals at $T_c$, which 
is approximately 10 K in a field of 11.5 T. Although this result implies 
that the same type of SC state sets in at all lattice sites, we caution 
that the drop in $1/^{23}T_1T$ is not at all sharp and the superconductivity
is weak at best. We find only a very modest decrease in $1/^{23}T_1T$, by a
factor of two from 10 K down to 2 K, whereas the data at $x = 0.019$, which
we present in Sec.~IV, show a much larger drop of $1/^{23}T_1T$ below $T_c$
for the PM phase. For completeness, we comment here that we also did not find 
an appreciable decrease in the Knight shifts ($^{23}K$ or $^{75}K$) below $T_c$ 
in the {\it w}-AFM phase (data not shown). However, it is worth noting that 
NMR results showing similar drops in $1/T_1T$ far below $T_N$ for undoped 
NaFeAs and CaFe$_2$As$_2$ systems have been interpreted as a type of activated 
behavior of magnetic domain-wall motion \cite{Ma_NaFeAs, Curro2}. We leave 
to a future study the investigation of whether the {\it w}-AFM phase may in 
fact arise from magnetic domain walls, whose characteristic width gives the 
small but doping-insensitive volume fraction we observe. 

In this regime we can only report that our current data are not sufficient
to differentiate between a scenario of microscopic coexistence, which would
be expected to show a far clearer signal, and a scenario where the PM regime
is a domain-boundary phase whose SC coherence length, unlike 245, exceeds
the domain size (the length scale of the nanoscopic phase separation), causing 
proximity superconductivity to pervade the entire magnetic regime. We can state 
that our results are fully consistent with STM data for a similarly underdoped
system ($x =$ 0.014) \cite{WangYY_NC_2012}, which show phase inhomogeneity, a
SC gap on all sites, and a strong anticorrelation (competition) between the 
AFM and SC order, a topic we discuss next (Sec.~IV). To summarize our analysis 
of the underdoped regime, our data for the constant {\it w}-AFM volume fraction 
and the common magnetic onset temperatures for {\it s-} and {\it w}-AFM are 
strong evidence in support of proximity magnetism in the {\it w}-AFM/PM phase. 
We return in Sec.~VI to a detailed discussion of the phase separation between 
the PM and {\it s}-AFM regions, and of its implications for iron-based 
superconductivity.

\begin{figure}
\includegraphics[width=8.5cm]{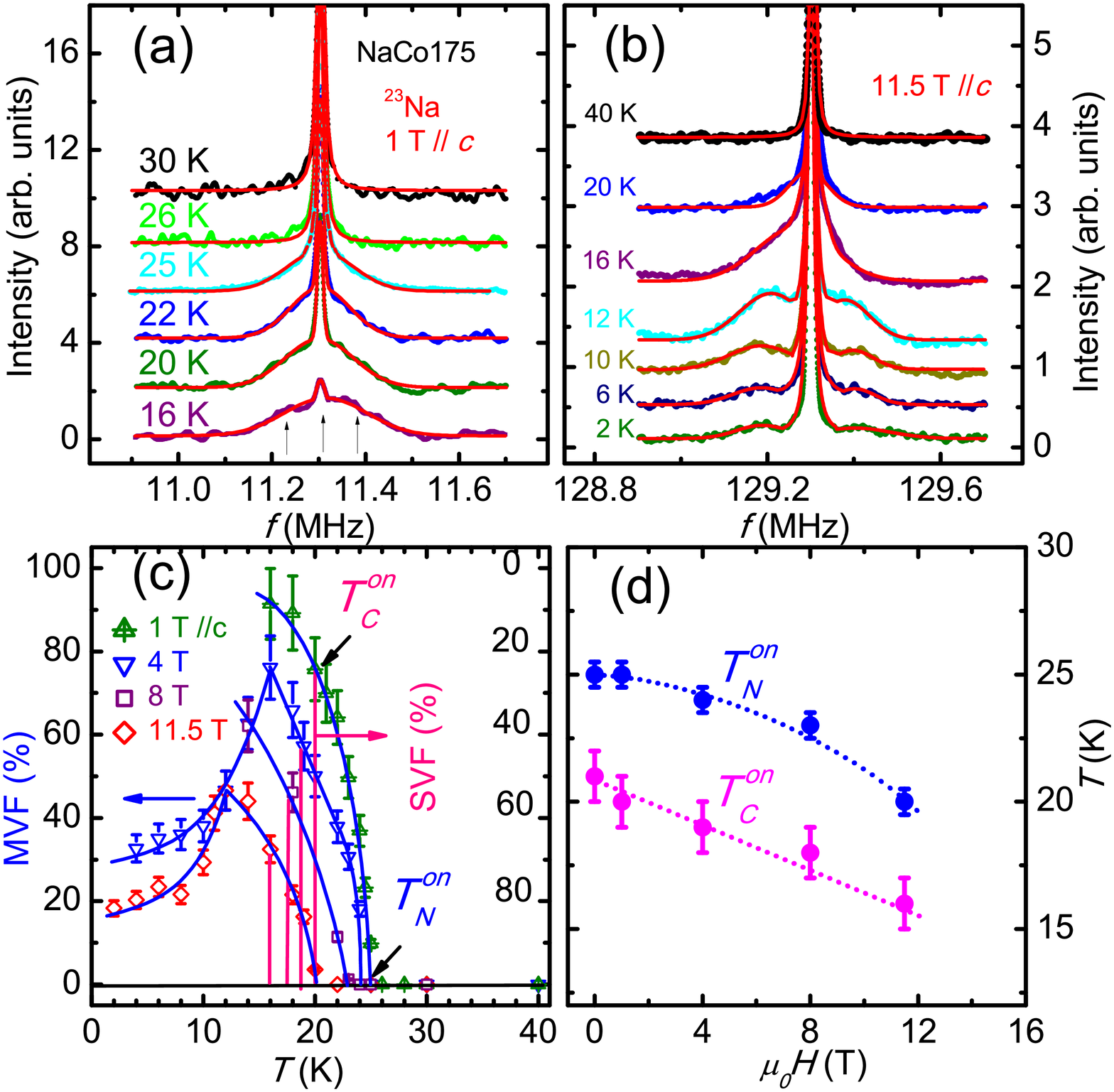}
\caption{\label{spec2}(Color online) $^{23}$Na central line for the NaCo175
crystal at selected temperatures in fields of (a) 1 T and (b) 11.5 T applied
along the $c$-axis. Solid curves are fits to the spectrum with one or more
Gaussian functions. (c) Temperature dependence of the magnetic (MVF) and
superconducting (SVF) volume fractions at each field, deduced from the
Gaussian fits. (d) Field dependence of the onset temperatures of static
antiferromagnetism, $T^{on}_N$, and superconductivity, $T^{on}_c$. }
\end{figure}

\section{Field-Controlled Volume Competition}

We turn next to our results for crystals with slightly higher doping levels,
$x = 0.0175$ (which we label NaCo175) and $x = 0.0190$ (NaCo190). The NaCo175
sample has a lower onset N{\'e}el temperature and smaller ordered moment,
while NaCo190 shows no long-ranged magnetic order, and so these samples
represent the evolution of the system to optimal doping.

Figure \ref{spec2}(a) shows the $^{23}$Na spectra at selected temperatures
in a field of 1 T oriented along the crystal $c$-axis. The spectrum at 30 K
shows a narrow line with a FWHM of approximately 5 kHz. On cooling to 25 K,
a shoulder feature develops at both sides of the peak, which on further
cooling grows in weight, whereas the weight of the center peak decreases.
This feature indicates the development of two phases at low temperature,
with PM sites giving the sharp center peak and magnetic sites giving the
broad shoulders, which in contrast to undoped NaFeAs \cite{Ma_NaFeAs} show
a wide distribution of local fields. Here $T^{on}_N =$ 25 K is the onset
temperature of antiferromagnetism. By measuring the RF inductance, we find 
that the onset temperature of superconductivity is $T^{on}_c =$ 20 K, and 
below 16 K the spectrum becomes too small to detect because of strong RF 
screening in the SC phase.

As shown in Fig.~\ref{spec2}(a), the spectra below $T_N$ can be fitted by
two superposed Gaussian functions. We deduce the magnetic volume fraction
(MVF) from the ratio of the spectral weight, taken from the Gaussian fit,
of the magnetic (shoulder) feature to the total weight. The MVF at a field
of 1 T is shown as a function of temperature in Fig.~\ref{spec2}(c), where
it clearly starts to develop at $T^{on}_N =$ 25 K, and increases with cooling.
At $T = 16$ K, the MVF reaches 90$\%$, indicating that the sample is almost
entirely magnetic. The average ordered moment for the magnetic part can also
be estimated from the NMR spectrum, as the extension of the shoulder away
from the central peak reflects the increase of internal static field (ordered
moment). At $T = 16$ K, the FWHM of the magnetic part of the spectrum is
approximately 150 kHz, which corresponds to $28\%$ of the moment in the
parent compound NaFeAs \cite{Ma_NaFeAs}, or in other words an average moment
of $\mu \simeq 0.09 \mu_B$/Fe with a spatially inhomogeneous distribution.
This behavior suggests that AFM order develops in islands below $T^{on}_N$ and
enlarges on cooling both in moment size and especially in volume fraction.

At higher magnetic fields, both $T^{on}_N$ and the MVF are suppressed. Figure
\ref{spec2}(b) shows $^{23}$Na spectra in a field of 11.5 T applied along
${\hat c}$. The spectrum is single-peaked and sharp above 20 K, with no
shoulder feature and hence no static magnetism. Below this, the shoulder
appears and we fit the spectrum with three Gaussian functions to account for
the center (PM) and shoulder (magnetic) components. In Fig.~\ref{spec2}(c),
the MVF at 11.5 T is seen to increase on cooling, similar to the low-field
data, but with a lower onset temperature ($T^{on}_N =$ 20 K) and a lower
MVF $\simeq 50\%$ at 12 K. Similar results for intermediate fields, also
shown in Fig.~\ref{spec2}(c), demonstrate the continuous nature of these
effects.

The most striking feature of Fig.~\ref{spec2}(c) occurs at the onset 
of superconductivity ($T^{on}_c$). Above $T^{on}_c$, the MVF increases 
monotonically on cooling at a fixed field, but below $T_c$ it falls
away; at 11.5 T ($T_c$ = 18 K), the MVF decreases from 50$\%$ at 12 K to
18$\%$ at 2 K. This behavior demonstrates a direct competition for volume
fraction between antiferromagnetism and superconductivity, which is also 
visible in the spectra shown in Fig.~\ref{spec2}(b). In Fig.~\ref{spec2}(d), 
we show both $T^{on}_N$ and $T^{on}_c$ as functions of field. $T_N$ shows a 
quite significant decrease with field, which can be fitted by the functional 
form $T_N = T_N (0) \sqrt{1 - (H/H_c)^2}$, producing an estimate of the 
critical field for $T^{on}_N = 0$ to occur at $H_c \approx 19 \pm 1$ T. By 
contrast, the field-induced decrease of $T^{on}_c$ is slower, consistent with 
the critical field for superconductivity being located at $H_{c2} \approx 50$ 
T in this system \cite{Wright_PRB_2014}, and suggesting that antiferromagnetism 
is suppressed at far lower fields than superconductivity. We comment here that 
such strong field effects on $T_N$ are highly unusual in iron-based SC 
materials, where the in-plane magnetic interactions are normally many tens
of meV, and we stress that this result is obtained only for our samples
close to optimal doping. For the underdoped samples discussed in Sec.~III,
we found no significant field-induced changes either to $T_N$ or to the
{\it s}-AFM volume fraction up to 12 T (data not shown). We return to
this issue in Sec.~VI.

Indeed our MVF data demonstrate that superconductivity is more stable than 
antiferromagnetism, replacing it at low temperatures for all fields. We will 
show later that for $x = 0.0175$, superconductivity occupies the PM phase, 
but not the AFM phase, during the replacement process. If the RF screening 
is non-uniform and strong in the SC regions, then our data provide an upper 
bound for the MVF below $T_c$. The drop of MVF on cooling indicates that the 
SC volume fraction, meaning the fraction of a percolating SC state, increases. 
However, the average moment in the magnetic regions remains large, $\mu \simeq 
0.09 \mu_B$/Fe, when $T \ll T_c$, and therefore we observe that AFM and SC 
order compete over the system volume, excluding each other in a first-order 
manner rather than coexisting with a reduced order parameter. Although 
competitive behavior of antiferromagnetism and superconductivty has been 
reported by neutron scattering studies of the Ba(Fe$_{1-x}$Co$_x$)$_2$As$_2$ 
system \cite{Fernandes}, these cannot distinguish whether the phases compete by 
order-parameter suppression (second order) or volume suppression (first order). 
Our unambiguous demonstration of volume competition in NaFe$_{1-x}$Co$_x$As is 
a key result whose origin and implications we discuss in Sec.~VI.

\begin{figure}
\includegraphics[width=8.5cm,height=6cm]{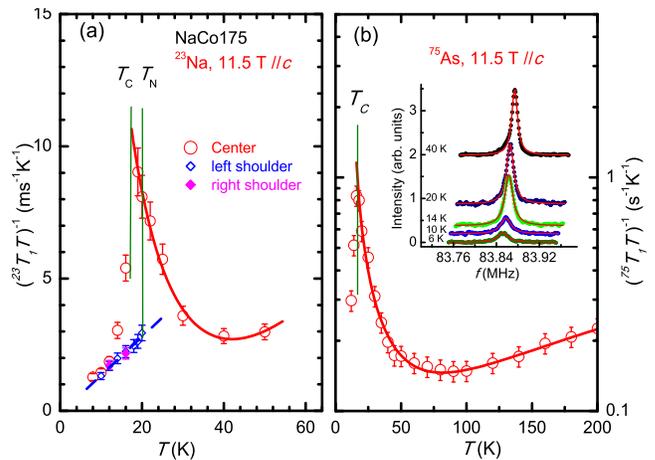}
\caption{\label{invt1t3}(Color online) (a) Spin-lattice relaxation rate
$1/^{23}T_1T$ measured on $^{23}$Na for NaCo175, shown as a function of
temperature with a field of 11.5 T applied along ${\hat c}$. Dotted lines
denote $T_N$ and $T_c$, circles are data measured at the peak of the spectrum
(PM phase), and diamonds are measured in the broad shoulder (AFM phase). Solid
lines are guides to the eye. (b) $1/^{75}T_1T$ measured on $^{75}$As for NaCo175
as a function of temperature with the same field. Inset: $^{75}$As NMR spectra
at selected temperatures. Solid lines are Gaussian fits to the data. }
\end{figure}

Turning now to further details of magnetism in the NaCo175 sample, we note
that the observed shoulder spectrum of $^{23}$Na is consistent with the
{\it s}-AFM phase reported in Sec.~III, but with the lower $T_N$ value
expected at a higher doping. However, there is no longer any evidence for
ordered moments in the $^{23}$Na PM signal, suggesting that the {\it w}-AFM 
phase is absent at this doping. We confirm that the absence of ordered 
moments is not a resolution issue by comparing with the $^{75}$As NMR
spectrum. As noted in Sec.~II, $^{75}$As has a much stronger hyperfine 
coupling than $^{23}$Na, and therefore its strong sensitivity to any weak 
magnetic order makes it the optimal probe for excluding a {\it w}-AFM 
component in the PM signal. As shown in the inset of Fig.~\ref{invt1t3}(b), 
the $^{75}$As spectrum is single-peaked at all temperatures. A narrow line 
with FWHM $\simeq$ 14 kHz is observed above $T^{on}_N = 20$ K, and below this 
its spectral weight begins to decrease due to the increasing AFM volume 
fraction. The AFM signal lies outside our $^{75}$As measurement window, 
because its FWHM is very large (it can be estimated from the $^{23}$Na 
data to be around 2.4 MHz). The single-peaked form of the $^{75}$As spectrum 
below $T^{on}_N$ is consistent with the PM $^{23}$Na signal, which remains 
sharp on cooling. Below $T^{on}_c = 16$ K, the spectra shift downward, as 
expected for singlet superconductivity, and at this point the $^{23}$Na 
spectrum does become broadened; at $T =$ 6 K, far below $T_c$, we observe 
a FWHM $\simeq$ 24 kHz as a consequence of the vortex structure in the SC 
phase. This value of the FWHM sets a strict limit on the ordered moment of 
the PM phase, which should be less than $0.6\%$ of the moment in NaFeAs 
(0.32$\mu_B$/Fe), and thus effectively excludes any possibility of 
{\it w}-AFM character at $x = 0.0175$.

Next we focus on the SC state of the NaCo175 sample. At low temperatures,
the PM phase is found to be purely superconducting by inspection of the
spin-lattice relaxation rates for both $^{23}$Na and $^{75}$As. Figure
\ref{invt1t3}(a) shows $1/^{23}T_1T$ in a field applied along the $c$-axis;
above $T^{on}_N$ we observe a decrease on cooling down to 40 K, followed by
an increase on further cooling below 40 K. The high-temperature behavior
is consistent with local-moment fluctuations \cite{Ma_prb_84} and the
low-temperature upturn with the spin fluctuations of itinerant electrons
at the Fermi surface \cite{Ma_prb_84, Ji_prl_2013}. Below $T^{on}_N$,
$1/^{23}T_1T$ is no longer uniform, showing a different form if taken at
different parts of the spectrum. For the shoulder, $1/^{23}T_1T$ drops quickly 
to a small value below $T^{on}_N$, as a consequence of the onset of static AFM 
order. For the peak, $1/^{23}T_1T$ continues to increase on cooling, falling 
only when the SC state is reached. Figure \ref{invt1t3}(b) shows the 
corresponding results taken from $^{75}$As, which are naturally uniform
because the spectra [inset, Fig.~\ref{invt1t3}(b)] have only a PM peak and
no shoulders. From $T = 220$ K down to 80 K, $1/^{75}T_1T$ decreases linearly
with temperature due to thermal excitation of local spin fluctuations in 2D
\cite{Ma_prb_84}. Below 80 K, the relaxation rate increases strongly with the
$1/^{75}T_1T = A/(T - \Theta)$ form characteristic of low-energy itinerant
spin fluctuations \cite{J.Mag.Mag.Mat.100}. The uniform sharp drop of $1/T_1T$ 
at $T_c$ for the PM signal of both nuclei indicates that the PM state becomes 
fully SC and it is believed from ARPES measurements that a full gap opens at 
all points on the Fermi surface; however, we comment that NMR data have not 
been able to verify this second point directly (the apparent linear form of 
$1/^{23}T_1T$ visible around $T = T_c/2$ in Fig.~\ref{invt1t3}(a), which may 
be of extrinsic or intrinsic origin).

\begin{figure}
\includegraphics[width=8.5cm, height=6cm]{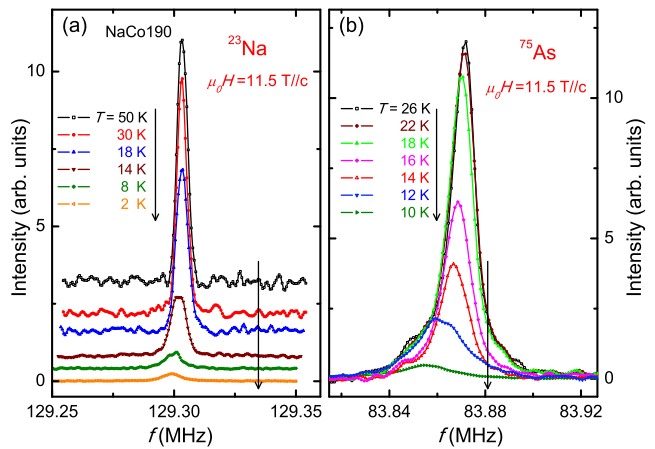}
\caption{\label{spec4}(Color online) (a) $^{23}$Na and (b) $^{75}$As NMR
spectra for NaCo190, shown at selected temperatures a field of 11.5 T
applied along ${\hat c}$.}
\end{figure}

The spin-lattice relaxation rates in Fig.~\ref{invt1t3}(a) suggest further
that the {\it s}-AFM phase is not strongly coupled to superconductivity.
Measurements of $1/^{23}T_1T$ for magnetic sites (in the shoulder of the
spectrum) show a drop at $T^{on}_N$, where magnetic order sets in, but there is 
no discernible drop at $T_c$. Thus there is no evidence that the {\it s}-AFM 
phase supports even weak or proximity superconductivity. We stress that
$1/^{23}T_1T$ at 12 K and 11.5 T, deep within the ordered phases, reaches
a similar value for sites in both the SC and the {\it s}-AFM regions. This
appears to be a clear statement that electrons on the Fermi surface are
gapped by either type of order, and, taken together with Fig.~\ref{spec2}(c),
that relaxation contributions become dominated by SC electrons at low
temperatures. This again reflects the fact that the competition between
antiferromagnetism and superconductivity for electrons on the Fermi surface 
(i.e.~in reciprocal space), and its apparent first-order nature, results in 
the volume competition (in real space) we observe in the NMR spectra.

Finally, we comment that further information concerning the volume-competition
effect can be gained by investigating samples with higher doping, namely $x
\simeq 0.019$. Our NaCo190 sample shows a structural transition at $T_s \simeq
35$ K, which we discuss in detail in Sec.~V. Figures \ref{spec4}(a) and (b)
show respectively the $^{23}$Na and the $^{75}$As spectra for different
temperatures. For $^{23}$Na, the spectrum is single-peaked and no shoulder
feature develops on cooling, even down to 2 K. This observation excludes
the existence not only of a possible {\it w}-AFM component but also of
the {\it s}-AFM phase. For $^{75}$As, the spectrum also has a single peak,
with FWHM $\simeq$ 40 kHz at the lowest temperatures, which also excludes
any type of AFM order. Below 18 K, the spectra shift to lower frequencies and 
a line broadening is clearly visible for $^{75}$As, which is the hallmark of
the onset of singlet-pairing superconductivity. Thus the effect of doping
on volume competition is to terminate the battle in favor of superconductivity 
at $x \simeq 0.019$, where one finds a single, uniform phase with only 
structural and SC transitions, but a complete absence of AFM order.

\section{Phase Diagram}

\begin{figure}
\includegraphics[width=8.5cm, height=8cm]{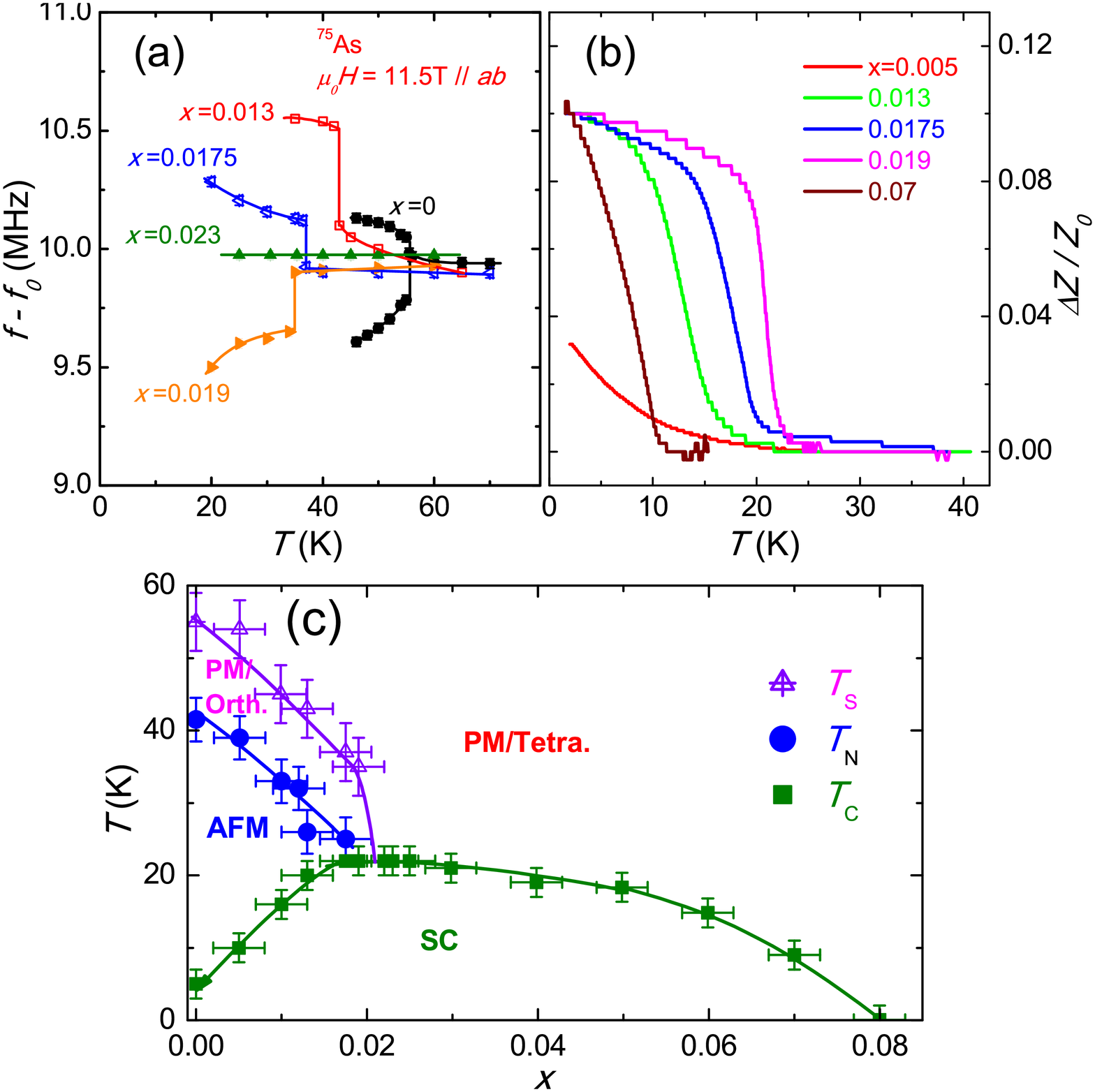}
\caption{\label{sus5}(Color online) (a) Determination of the structural
transition temperature $T_s$ from the frequency shift $f - f_0$ of the
satellite line in the $^{75}$As spectrum measured with an in-plane field,
shown for a range of sample dopings. (b) Determination of the SC onset 
temperature $T_c$ from measurements of the inductance change $\Delta Z/Z_0$ 
of the NMR coil at zero field, also for a range of dopings. (c) Phase
diagram showing the values of $T_N$ determined from the NMR spectra of
Figs.~\ref{tn1}(a) and \ref{tn1}(b), $T_s$ determined from the satellite
line shift, and $T_c$ determined from the RF inductance. }
\end{figure}

We now compile all of our results, from samples across the full range of
doping, to prepare a definitive $(x,T)$ phase diagram. First, the structural
transition can be detected from the frequency of a chosen satellite line in
the $^{75}$As spectrum \cite{Ma_NaFeAs}, which is shown in Fig.~\ref{sus5}(a)
as a frequency shift relative to the center line. When the field is applied
in the $ab$-plane, cooling from the tetragonal to the orthorhombic phase
causes each satellite to shift and to split into two due to sample twinning.
The sudden change of the satellite frequency as a function of temperature,
clearly visible in Fig.~\ref{sus5}(a), determines $T_s$ for the structural
transition at each doping. We comment here that above $T_s$ this satellite 
frequency, which for perfect field alignment is the quadrupole frequency 
$\nu_Q$, is generally expected to show a systematic increase with sample 
doping \cite{Lang_PRL_104_097001}; however, such a dependence is barely 
discernible in our data [Fig.~\ref{sus5}(a)] due to the fact that the maximum 
doping we studied in the NaFe$_{1-x}$Co$_x$As system is so low ($x = 0.023$). 
The variation in our measured values of $f - f_0$ above $T_s$ for the different 
samples is in fact dominated by the small but finite misalignment of the 
magnetic field, whose exact orientation with respect to the crystalline 
$a$-and $b$-axes was not determined. NMR provides an accurate measurement 
of $T_s$ up to $x = 0.019$, beyond which the sample remains tetragonal at 
all temperatures.

Next, the transition to static magnetic order can be detected at all dopings by 
the decrease of the PM spectral weight, as shown in Figs.~\ref{tn1}(a) and (b).
Finally, the SC transition is detected by the relative inductance change of 
the NMR coil, as noted in Sec.~II and shown in Fig.~\ref{sus5}(b); these 
measurements can also be benchmarked from the drop of the NMR Knight shift 
and the spin-lattice relaxation rate, at least for $x \ge 0.0175$ (see 
Sec.~III). In Fig.~\ref{sus5}(c) we present the hierarchy of deduced 
transition temperatures, which show the clear evolution characteristic 
of iron-based superconductors. In the parent compound, NaFeAs, the system 
becomes fully magnetic below $T_N \simeq 41.5$ K, but as electron doping 
is induced by Co substitution, the antiferromagnetism is gradually suppressed 
and superconductivity develops.

\begin{figure}
\includegraphics[width=8.5cm, height=11cm]{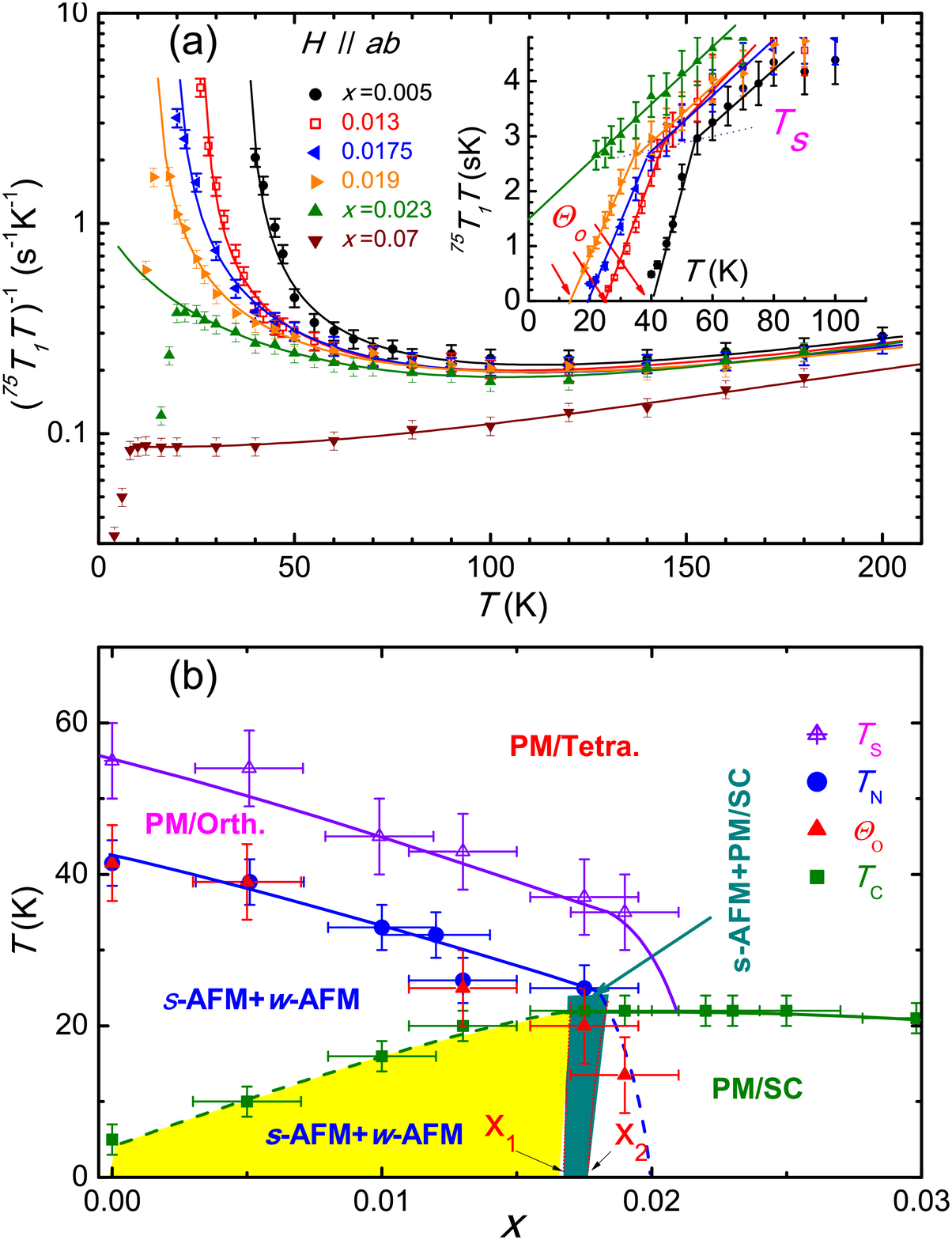}
\caption{\label{pd6}(Color online) (a) Temperature dependence of
$1/^{75}T_1T$ for NaFe$_{1-x}$Co$_x$As crystals under a field of 11.5 T
applied in the $ab$-plane of the crystal. Solid lines are fits to the
function $1/^{75}T_1T = A/(T - \Theta) + BT + CT^2$. Inset: $^{75}T_1T$
as a function of temperature. $T_s$ and $\Theta_o$ denote respectively
the temperatures where $^{75}T_1T$ changes slope and where it goes to
zero. (b) Phase diagram of NaFe$_{1-x}$Co$_x$As established by NMR: $T_s$, 
$T_c$, $\Theta_o$, and $T_N$ denote respectively the structural transition, 
the SC transition temperature, and the Curie-Weiss temperature, all obtained 
from the fit to $1/^{75}T_1T$ above $T_c$, and the N\'eel temperature obtained 
from the PM spectral weight [Figs.~\ref{tn1}(a) and (b)]. Data for $T_c$ below 
$x = 0.0175$ are taken only from inductance measurements [Fig.~\ref{sus5}(b)]. 
$x_1$ denotes the approximate doping where the {\it s}-AFM+{\it w}-AFM region 
and the {\it s}-AFM+PM region meet and $x_2$ denotes the doping where $T_N = 
0$ and beyond which AFM order is absent.}
\end{figure}

However, in NaFe$_{1-x}$Co$_x$As there is a complementary approach to obtaining
the phase diagram, which is that all three phase transitions can be detected
directly and accurately in the spin-lattice relaxation rate. This is not
possible in other pnictide materials, and as we explain below it is also a
consequence of the strongly 2D fluctuations acting in the 111 system. In
Fig.~\ref{pd6}(a) we show $1/^{75}T_1T$, obtained over the full range of
available dopings with the field applied in the $ab$-plane, a geometry
known \cite{Kita_JPSJ_77_114709,Ning_PRL_104} to maximize the sensitivity
of the measurement to the dominant in-plane spin fluctuations in pnictides 
as the $(\pi,0)$ magnetically ordered state is approached. At high 
temperatures, the $1/^{75}T_1T$ values are similar for all dopings other 
than the heavily doped $x = 0.07$ sample, and fall slowly on cooling down 
to 100 K. Below 100 K, a Curie-type upturn develops for all dopings (other 
than $x = 0.07$), becoming progressively stronger for lower doping values. 
Clearly the low-energy spin fluctuations responsible for this behavior, 
which have an itinerant origin, are suppressed strongly by increasing 
doping, reflecting a very high sensitivity to details of the Fermi surface. By 
contrast, the very small change above 100 K suggests an origin in local-moment 
(valence-electron) fluctuations, consistent with a general two-component 
interpretation of the spin response in pnictides \cite{Ma_prb_84}.

If the low-temperature $1/^{75}T_1T$ data are fitted with the function
$1/^{75}T_1T = A/(T - \Theta)$, one may extract a Curie temperature
$\Theta$. The structural transition can be deduced by considering the
quantity $^{75}T_1T$, as shown in the inset of Fig.~\ref{pd6}(a) for
different underdoped and near-optimal dopings. To a good approximation,
$^{75}T_1T = (T - \Theta)/A$ can be fitted with two straight lines of
different gradients, one (which we denote $\Theta_t$) corresponding to
the tetragonal structure and the other, $\Theta_o$, to the orthorhombic
one. The change in this gradient gives the structural transition temperature
$T_s$. The values of $\Theta_t$ and $\Theta_o$ can be obtained from the
intercept of the fitting line with $^{75}T_1T = 0$.

We find that the values of $\Theta_t$ (obtained by extrapolation from above
$T_s$) are all negative and decrease with doping, but these are of limited
physical meaning because the tetragonal structure is replaced at $T_s$,
perhaps precisely because it does not allow the magnetic transition from
which the orthorhombic phase can profit. By contrast, $\Theta_o$, which is
determined from the structure and data below $T_s$, is positive and decreases
with doping up to $x = 0.019$. The fact that $\Theta_o$ is far larger than
$\Theta_t$ reinforces the evidence that the low-energy spin fluctuations are
strongly enhanced below $T_s$, suggesting a clear role for the magnetic sector
in driving the structural transition. Although there is already significant
evidence for a coupling between the lattice structure and the magnetism
in 111 materials \cite{Ma_NaFeAs}, this coupling is manifestly not strong
enough that $T_N$ coincides with $T_s$, as it does in the 122 system. As
noted in Sec.~I, some authors \cite{XuCK} have taken the lack of coincidence
between $T_N$ and $T_s$ as a hallmark of strong two-dimensionality in some 
iron pnictide materials.

As noted above, the N{\'e}el temperature $T_N$ is determined directly from
the decrease of the PM spectral weight [Figs.~\ref{tn1}(a), \ref{tn1}(b), 
\ref{sus5}(c)]. For NaFe$_{1-x}$Co$_x$As, the fitting parameter $\Theta_o$, 
which we obtain from our data below $T_s$, has exactly the same value as 
$T_N$ for $x < 0.0175$. This extremely significant result has not been 
reported in any other iron pnictide systems. $T_N$ is a true measure of
static AFM order, setting in due to all of the couplings in the system. By 
contrast, $\Theta_o$ is a measure of the 2D spin fluctuations of the FeAs 
planes. For quasi-2D systems, the energy scale of in-plane fluctuations ($J$) 
sets a characteristic temperature $T_{\rm BKT}$ in the spin response, but true
long-range order is forbidden by the Mermin-Wagner theorem. However, a
weak coupling $J_c$ between these planes is sufficient to create long-range
order in three dimensions and the transition temperature will be $T_N \approx
T_{\rm BKT}$. Only when $J_c$ becomes a significant fraction of $J$ will the
planar physics be supplemented by conventional three-dimensional fluctuations
and $T_N$ will exceed $T_{\rm BKT}$ by an amount depending on $J_c$. The
magnetic interactions in the parent compound NaFeAs have been measured
directly in a very recent study of spin-wave dispersion relations by
inelastic neutron scattering \cite{Zhang_PRL_112_217202}. These authors
find in-plane couplings $SJ_{1a} \simeq 40$ meV, $SJ_{1b} \simeq 16$ meV,
and $SJ_2 \simeq 19$ meV, respectively for superexchange processes in the
$a$, $b$, and diagonal ($a \pm b$) directions in the FeAs plane, where $S$
is the effective spin (moment) on the Fe ions, but an interplanar coupling
of only $SJ_c \simeq 1.8$ meV. Thus our observation of perfect coincidence
between $\Theta_o$ and $T_N$ is completely consistent with the measured 
value $J_c/J \sim 0.1$, reflecting the minimal contributions from interplanar
coupling, and these results form the best evidence yet available for the
extremely 2D nature of the NaFeAs system. We defer a discussion of the
microscopic implications of this result to Sec.~VI.

We are now in a position to present the complete phase diagram, which
is shown in Fig.~\ref{pd6}(b). We begin by drawing attention to the error
bars on the doping ($x$) axis, which given the extreme sensitivity of the
system to small changes in $x$ is critical information. As noted in Sec.~II,
we do not possess probes capable of determining the doping level to the
0.1\% accuracies mandated by the physics of NaFe$_{1-x}$Co$_x$As, and so
we have taken the nominal doping in every case. However, we are able to
benchmark our samples against each other and the remarkably smooth evolution
in properties, as well as the perfect match between the $T_s$ datasets, shown
in Figs.~\ref{sus5}(c) and \ref{pd6}(b), indicates that our conservative
estimate of the doping error as $\pm$ 0.2\% is reasonable. This smooth
evolution over such a fine range of doping also supports the suggestion that
NaFe$_{1-x}$Co$_x$As provides some of the most homogeneous and highest-quality
crystals of any iron-based SC material. Returning to the matching of datasets 
between Figs.~\ref{sus5}(c) and \ref{pd6}(b), the agreement in $T_s$ benchmarks 
the value of the analysis of Fig.~\ref{pd6}(a); the agreement in $T_c$ is 
perfect for $x \ge 0.0175$, but the values extracted from $1/^{75}T_1T$ at 
lower dopings are inaccurate for the reasons discussed in Sec.~III, and so 
the data shown in Fig.~\ref{pd6}(b) for the underdoped samples are those of 
Fig.~\ref{sus5}(c); the agreement between $T_N$ and $\Theta_o$ is remarkable 
for the reasons discussed above.

Our NMR data [Figs.~\ref{sus5}(c) and \ref{pd6}(b)] provide direct
evidence for a complete separation of $T_s$ and $T_N$ at all dopings in
NaFe$_{1-x}$Co$_x$As, with the structural transition always occurring some
12$-$15 K above the magnetic one. Thus, in common with the majority of 
iron-based SC systems, the coupling between the lattice and the magnetism 
is ``subcritical'' in the sense of a combined transition. The $T_s$ line 
terminates at a slightly higher doping than $T_N$, and in fact at a value 
coinciding with optimal doping, as indicated also in transport studies 
\cite{Chen_NaFeCoAs}. The $T_c$ line is quite flat in the doping range 
$0.017 < x < 0.022$, whereas all of $T_s$, $\Theta_o$, and $T_N$ change 
dramatically. In the zero-temperature limit, antiferromagnetism and 
superconductivity cohabit for $0 < x < 0.0175$, as shown in Sec.~IV 
[Fig.~\ref{spec2}(c)], but within the phase-separated {\it s}-AFM and PM 
({\it w}-AFM) regions (Sec.~III). We comment again here that the onset of 
the {\it w}-AFM phase at the same temperature, $T_N$, as the {\it s}-AFM 
one suggests that the former is actually a set of narrow PM regions ordered 
by a proximity effect. Above $T_c$, the {\it s}-AFM and {\it w}-AFM phases 
``coexist,'' by which we mean ``cohabit as phase-separated regions,'' in 
the doping range $x < x_1$ and the {\it s}-AFM and PM phases cohabit over 
the range $x_1 < x < x_2$, as represented in Fig.~\ref{pd6}(b). Our data 
fix the values $x_1 \simeq 0.0175$ and $x_2 \simeq 0.018$.

To determine the detailed structure of the phase diagram close to optimal
doping, we show in Fig.~\ref{pd6}(b) all of the values of $\Theta_o$
determined from the Curie-Weiss fit below $T_s$. As noted above, $\Theta_o$
coincides with $T_N$ for $x \le$ 0.0175. However, for doping $x = 0.019$,
$\Theta_o$ determined from the $1/^{75}T_1T$ curve above $T_c$ predicts a
finite N{\'e}el temperature (13.5 K) below $T_c$ (22 K). In a scenario where
antiferromagnetism and superconductivity coexist, one might expect a magnetic 
quantum critical point ($\Theta_o = 0$) to occur at $x \approx$ 0.02. However, 
we demonstrated at the end of Sec.~IV that AFM order is completely absent for 
$x = 0.019$. Thus we conclude that in NaFe$_{1-x}$Co$_x$As, the appropriate 
scenario for the regime $T_c > \Theta_o$ is a complete suppression of 
long-range AFM order by volume competition (Sec.~IV).

Given that $\Theta_o$ $(T_N) =$ 25 K at $x = 0.0175$ and $\Theta_o =$ 13.5 K
at $x = 0.019$, it is reasonable to estimate that the AFM transition line
touches the SC dome at $x \approx$ 0.018, although we do not have samples 
with this doping. Because superconductivity suppresses the MVF below $T_c$, 
we expect that the {\it s}-AFM transition line is pushed to lower dopings
below $T_c$, as also represented in the slope of the line to ($x_2,0$) in
Fig.~\ref{pd6}. For all dopings ($x$) beyond this line, our experiments
show unequivocally that the system is single-phased with only one
transition, the onset of superconductivity in a structurally tetragonal
system.

\section{Discussion}

Our NMR measurements across the phase diagram in NaFe$_{1-x}$Co$_x$As reveal
a number of key features, including the first-order volume competition, the
microscopic phase separation, and the dominance of superconductivity at low
temperatures. We discuss these points in turn, finding that their common
denominator is the two-dimensional nature of the NaFeAs system.

Beginning with the volume competition effect, the large changes of the MVF we 
observe (Sec.~IV) as functions of both temperature and field suggest a mutual 
exclusion of AFM and SC order in real space. Such a real-space competition 
implies a first-order transition between two states with finite order 
parameters, which would appear to reflect a strong competition in reciprocal
space, meaning for electrons at the Fermi surface. Certainly the fact that 
the low-energy spin fluctuations, which are due to itinerant (Fermi-surface)
contributions \cite{Ji_prl_2013}, are fully gapped below both $T_N$ and $T_c$
for the $x =$ 0.0175 sample suggests that both types of order, stabilized by
their own particular Fermi-surface electronic order parameter ($\langle
c_{\bf k \uparrow}^{\dag} c_{{\bf k} + {\bf Q} \downarrow} \rangle$ or $\langle
c_{{\bf k}\uparrow}^{\dag} c_{- {\bf k} \downarrow}^{\dag} \rangle$), compete for the
same electrons.

Such a temperature- and field-controlled magnetic and SC volume fraction 
has not been reported in any other iron pnictides. We suggest that there are
two reasons why this highly unconventional phenomenon has been observed (to
date) only in NaFe$_{1-x}$Co$_x$As. One is the extreme 2D nature of the Fermi
surfaces and the other is the very fine control of the doping level, which
is obtained in NaFeAs systems with no loss of chemical homogeneity.

Addressing first the two-dimensionality, the question of whether 
electronic correlation effects result in competition or coexistence between 
antiferromagnetism and superconductivity has been fraught with contradictions 
in the iron-based SC systems. Microscopic coexistence of AFM and SC order has 
been reported in several compounds with the 122 structure, including by some 
of us \cite{Laplace_prb, Ma_prl_109_197002, Li_prb_86_180501, Iye_JPSJ, 
Julien_epl_2009}. AFM and SC states in iron-based SC materials depend rather 
sensitively on the interactions of quasiparticles at the Fermi surface, and 
as a result it was suggested in Ref.~\cite{Ma_prl_109_197002} that the ability 
of an iron-based SC system to host both types of order may be dictated by the 
variety of projected 2D Fermi surfaces available. The $c$-axis band dispersion 
in BaFe$_2$As$_2$ is quite significant \cite{Vilmercati_PRB_79_220503}, making 
it possible that the 122 structure may allow the coexistence of the two types 
of order for electrons on different parts of the Fermi surface. In 
Ba(Fe$_{1-x}$Ru$_x$)$_2$As$_2$, the fact that $1/^{75}T_1T$ drops due to gap 
formation at $T_N$ and again at $T_c$ demonstrates that additional SC electrons 
are present in the AFM phase \cite{Ma_prl_109_197002}. By contrast, this 
dispersion is much weaker in the NaFe$_{1-x}$Co$_x$As system, as shown both 
by ARPES studies \cite{Feng_JPCS_2011, WangSC_APL_2012} and by the present 
results (Sec.~V), indicating a highly 2D system with little flexibility in 
Fermi-surface sizes and spanning wave vectors. This leaves little option for 
the AFM and SC order but to ``fight it out'' for the available electrons, 
leading to the strong competition we observe (Sec.~IV): in NaFe$_{1-x}$Co$_x$As, 
$1/^{23}T_1T$ drops only at $T_N$ or at $T_c$, and to the same low-temperature 
value, showing directly that antiferromagnetism and superconductivity compete 
for the same electrons. Our results suggest strongly that the microscopic 
coexistence of AFM and SC order is not possible in NaFe$_{1-x}$Co$_x$As, which 
among other things should exclude any intrinsic superconductivity on sites 
with {\it s}-AFM order in the underdoped region (Sec.~III).

On this note, we turn next to the issue of phase separation, for which we 
find evidence throughout the phase diagram. Clearly distinguishable {\it s}-AFM 
and {\it w}-AFM regions are present in the underdoped system (Sec.~III), 
although the volume fraction of the {\it w}-AFM phase barely changes with 
either doping or temperature and there is only one magnetic transition.
At near-optimal underdopings, there is clear phase separation between the
AFM and PM/SC regions, demonstrated very explicitly by the large changes of
AFM and SC volume fractions we observe in our NaCo175 sample as functions of
both temperature and field (Sec.~IV), which strongly suggest a first-order
transition between the two phases. The phase diagram of Fig.~\ref{pd6}(b)
shows directly that the optimal $T_c$ is achieved when both the AFM and
orthorhombic phases are suppressed. Although this type of phase diagram
has been interpreted as an incipient AFM quantum critical point of the
orthorhombic system, in reality its suppression by the onset of the
mutually exclusive (volume-competitive) SC phase is the dominant physics.

Such a complex phase diagram may also exist in other iron-based SC systems,
although, for reasons of the fine doping control mentioned above, none has 
yet been resolved in the kind of detail possible in NaFe$_{1-x}$Co$_x$As. 
Within this intricate phase structure, the inevitable presence of weak 
disorder could certainly provide an extrinsic origin for the phase separation 
of {\it s}-AFM and {\it w}-AFM regions in the underdoped regime. However, we 
appeal again to the evidence for remarkably high sample homogeneity in 
NaFe$_{1-x}$Co$_x$As and to the obvious fact of very low carrier densities. 
If the origin of phase separation between the {\it s}-AFM and {\it w}-AFM 
regions is the same as that between the {\it s}-AFM and PM/SC regions near 
optimal doping, then such behavior may in fact be intrinsic in systems with 
strong electronic correlations. In cuprate materials, the stripe phase 
\cite{Tranquada} may be considered as an atomic-scale phase separation of 
AFM and PM/SC regions. Similar phase separation has been reported recently 
close to the phase boundary both in an organic superconductor 
\cite{Brooks_PRL_2014} and in heavy-fermion superconductors tuned 
by pressure and magnetic field \cite{Seo_NP_2014}.

Complete nanoscale phase separation is familiar in iron-based SC materials 
from the case of the 245 iron selenides. However, as noted in Sec.~III, the 
situation in NaFe$_{1-x}$Co$_x$As does not appear to be the same, first in that 
structural and hence doping inhomogeneity is significant in 245 materials and 
second in that weak superconductivity can be observed throughout the sample 
below $T_c$. In Sec.~III we outlined two scenarios for the present result, a 
microscopic coexistence or a microscopic phase separation, and, as noted 
above, the results of Sec.~IV make a coexistence appear extremely unlikely. 
Although we cannot claim evidence for a nanoscale phase separation from our 
data, all of our results are consistent with such a scenario, under the 
proviso that the length scale of the phase-separation phenomenon be extremely 
short. We have observed a phase susceptible to an apparent bulk magnetic order 
by proximity effects, which is a definite statement that the phase-separation 
length scale should be short compared to the magnetic correlation length. 
The weakness and feeble onset of the apparent bulk SC regime could be the 
fingerprints of proximity superconductivity originating in narrow PM regions 
but spreading throughout the very small magnetic domains. Microscopically, in 
the absence of doping inhomogeneity effects, the scale of the phenomenon is 
expected to result from a subtle interplay between electronic correlations 
and lattice or charge inhomogeneities (independent of Co doping), and could 
indeed be on the nanometer scale.

First-order phase separation close to the AFM quantum phase transition
has certainly been suggested in other iron-based SC materials, most notably
BaFe$_{2-x}$Ni$_x$As$_2$, although the fact that $T_s$ and $T_N$ merge in
the 122 systems may cause qualitative differences in the phase diagram.
One key proposal from these neutron scattering studies is an incommensurate
nature of the resulting AFM phase \cite{LuXY}. If the direct volume competition
of the AFM and SC phases we observe below $T^{on}_c$ takes place in a nanoscale
lamellar structure, then an incommensurate AFM signal may indeed be observed.
A different interpretation, namely a double quantum critical point, has been
offered from combined transport and NMR studies in BaFe$_{2-x}$Ni$_x$As$_2$
\cite{ZhengGQ}, but the magnetic structure below $T_c$ was not resolved. NMR
measurements on Ba(Fe$_{1-x}$Co$_x$)$_2$As$_2$ suggest a cluster spin-glass
phase close to optimal doping \cite{CurroNJ}; in NaFe$_{1-x}$Co$_x$As this
type of physics can be excluded explicitly from our data, which do not contain 
the stretched spin recovery of a spin glass, indicating again the high quality
of our samples.

One of the fundamental questions in formulating a microscopic model for 
the iron-based SC materials is the equitable treatment of local-moment 
(valence-electron) and itinerant (conduction-electron, or Fermi-surface) 
contributions to the macroscopic properties of magnetism and pairing. 
Returning again to the key observation (Sec.~IV) of large changes in the 
MVF with both temperature and field, these suggest a microscopic phase 
separation, which one expects to be driven by a first-order magnetic 
quantum phase transition. Because the field is applied along the $c$-axis, 
which is perpendicular to the direction of the ordered moments, a spin-flop 
transition cannot account for the observed field effect. Rather, the strong 
suppression of $T^{on}_N$ by an applied field near optimal doping (Sec.~III; 
the rate is approximately $- 0.5$ K/T close to $H = 0$), the classical 
critical behavior with increasing field [Fig.~\ref{spec2}(d)], and the 
extreme field sensitivity to the competing SC phase close to the putative 
magnetic quantum critical point, all suggest that the interactions causing 
AFM order are largely local in nature.

In this context it is worth remarking that for overdoped systems up
to $x =$ 0.07, where low-energy spin fluctuations are entirely absent
[Fig.~\ref{pd6}(a)] but $T_c$ is still approximately 10 K, it is local-moment
spin fluctuations originating in the valence electrons that provide the
pairing interactions for superconductivity. High-pressure NMR studies in
overdoped NaFe$_{1-x}$Co$_x$As samples demonstrate direct contributions to
superconductivity from both itinerant-electron spin fluctuations (which
are pressure-dependent) and local-moment fluctuations (which are largely
pressure-independent) \cite{Ma_prl_109_197002}. Thus it is clear that both
local-moment magnetism and itinerant electrons are required for a complete
understanding of the nature of antiferromagnetism and superconductivity, and 
in this light we revisit the key result of Sec.~V that $\Theta_o = T_N$. The 
low-energy spin fluctuations causing the Curie-Weiss upturn in $1/T_1T$, and 
hence determining $\Theta_o$, are due only to Fermi-surface electrons, so 
the equality with $T_N$ indicates only very weak contributions to long-range
order from an inter-plane interaction $J_c$ (Sec.~V), which from the previous 
paragraph we conclude is mediated by local-moment fluctuations. This result, 
demonstrating clearly the extremely 2D nature of the NaFeAs system, suggests 
that the cornerstone of a microscopic model should be the itinerant 
contribution to (planar) magnetic order, without which the local-moment 
interactions appear quite unable to order alone.

Finally, we comment that the suppression of antiferromagnetism by 
superconductivity at low temperatures for the $x = 0.0175$ sample, even 
though $T_N$ exceeds $T_c$, indicates that the SC phase possesses additional 
electronic or magnetic channels with which it can ``win'' against AFM order. 
A full microscopic model of the band structure, Fermi surfaces, and 
correlation effects due to valence-electron contributions is required to 
account for this effect, and here we can only present the possibilities 
suggested by our data. One may be that superconductivity competes only with 
the AFM tendencies of the itinerant electrons, which in other iron-based 
superconductors are reinforced by sizeable local-moment contributions, but 
that (previous paragraph) the extremely 2D nature of NaFeAs weakens this 
link to the point that the AFM phase is disrupted completely.

The microscopic physics underlying such behavior will be found naturally in
an orbital-specific model. Even in a fully 2D system, the electronic bands
of iron pnictides contain five different $d$-orbitals, and hence many degrees 
of freedom in orbital symmetries and admixtures. Antiferromagnetism and 
superconductivity are favored by itinerant electrons in bands of different 
orbital content, and the unique feature of NaFe$_{1-x}$Co$_x$As is that the 
band occupations, and thus their Fermi surfaces, are inordinately sensitive 
to the doping because of the highly 2D nature of the system. A recent ARPES 
study of NaFe$_{1-x}$Co$_x$As samples with small and large $x$ \cite{Ye_PRX_2013}
has provided some indications for orbital-selective connections between the 
competing AFM and SC phases, specifically concerning the relative $d_{xy}$ 
and $d_{xz}/d_{yz}$ content of the bands near the Fermi surface. We suggest 
that similarly detailed studies of samples near optimal doping have the 
potential to reveal the underlying physics of NaFeAs.

In summary, by using NMR as a local probe sensitive to both antiferromagnetism
and superconductivity, we observe a strong volume competition between the two
phases at the boundary of the antiferromagnetic phase transition in
NaFe$_{1-x}$Co$_x$As. The volume fractions of the two phases can be controlled
by varying both the temperature and the applied magnetic field, and show a
complete mutual exclusion in real space. Thus our NMR data support a
first-order phase transition between antiferromagnetism and superconductivity,
which is driven by the competition between their electronic order parameters
in reciprocal space. As striking as the volume competition effect is the
exquisite sensitivity of the competition to doping, with optimal doping
and all of the phase-separation effects occuring between 0 and 2\%. These
phenomena have their origin in the extremely weak interplane coupling in
the NaFeAs materials, resulting in a very two-dimensional nature of the
electronic band structure, and hence of the Fermi surfaces. One key
anomaly compared to other iron-based superconductors is the winning of
superconductivity over antiferromagnetism in real space at moderate
underdopings, even where the magnetic transition temperature is higher,
suggesting that the generic behavior of a two-dimensional iron pnictide may
be for the electronic or magnetic channels of the Fermi-surface electrons
to favor superconductivity. Further, because the very weak Co doping also
appears to be remarkably homogeneous, NaFe$_{1-x}$Co$_x$As is an excellent
system in which to seek evidence of unconventional phases arising purely
due to intrinsic electronic correlations. For this we obtain additional
information concerning the {\it w}-AFM minority phase, which may be a
paramagnetic regime occurring as thin lamellae due to nanoscale phase
separation, but appears antiferromagnetic by proximity for underdoped
samples and is the first region to turn superconducting on the approach 
to optimal doping.

\acknowledgments

We are grateful to T. Giamarchi, S. P. Kou, H. H. Wen, Z. Y. Weng, and R. Yu
for helpful discussions. Work at Renmin University of China was supported by
the National Basic Research Program of China under Grant Nos.~2010CB923004,
2011CBA00112, and 2012CB921704 and by the NSF of China under Grant
Nos.~11174365, 11222433, and 11374364. Work at Rice University 
and the University of Tennessee in Knoxville was supported by the
US DOE Office of Basic Energy Sciences through contracts DE-SC0012311 (P.D.) 
and DE-FG02-08ER46528 (C.L.Z).

\end{document}